\def\kms{\ifmmode {\rm \, km \, s^{-1}} \else $\rm \,km \, s^{-1}$\fi}
\def\etal{{\sl et al.\ }}

\documentstyle[11pt,aaspp4]{article}

\received{19 October 1995}
\accepted{7 November 1995}

\slugcomment{\it Accepted for publication in the Astrophysical Journal Letters}

\lefthead{Courteau et al.}
\righthead{Secular Evolution in Late-Type Spirals}

\begin{document}


\title{Evidence for Secular Evolution in Late-Type Spirals}

\author{St\'ephane Courteau\altaffilmark{1}}
\affil{Kitt Peak National Observatory, 950 N. Cherry Ave., Tucson, AZ 85726}
\affil{E-mail: courteau@noao.edu}

\author{Roelof S. de~Jong\altaffilmark{2}}
\affil{University of Durham, Dept.\ of Physics, South Rd., Durham, DH1 3LE,
England}
\affil{E-mail: R.S.deJong@durham.ac.uk}

\and

\author{Adrick H. Broeils\altaffilmark{3}}
\affil{Center for Radiophysics and Space Research, Cornell University, Ithaca,
  NY 14853}
\affil{E-mail: broeils@astro.su.se}



\altaffiltext{1}{Visiting Astronomer at Lick, Palomar,
 and Kitt Peak National Observatory. KPNO is operated by AURA, Inc.\ under
 cooperative agreement with the National Science Foundation.}

\altaffiltext{2}{Visiting Astronomer at Jacobus Kapteyn Telescope (La Palma)
 operated by the Royal Greenwich Observatory and the UK Infrared Telescope
 operated by the Royal Observatory Edinburgh with financial support of the
 PPARC.}

\altaffiltext{3}{Present address: Stockholm Observatory, S-133
  36 Saltsj{\"o}baden, Sweden}


\begin{abstract}
We combine deep optical and IR photometry for 326 spiral galaxies from two
recent galaxy samples and report that the surface brightness profiles of
late-type
spirals are best fit by two exponentials. Moreover, the ratio of bulge and disk
scale lengths takes on a restricted range of values and is uncorrelated with
Hubble type. This suggests a scale-free Hubble sequence for late-type spirals.
Careful numerical simulations ensure that our results are not affected by
seeing or resolution effects. Many of these galaxies show spiral structure
continuing into the central regions with a previously undetected small bar
and slowly changing colors between the inner disk and the bulge. We invoke
secular dynamical evolution and interpret the nature of disk central regions in
the context of gas inflow via angular-momentum transfer and viscous transport.
In this scenario, galaxy morphologies in late-type spirals are not imprinted at
birth
but are the result of evolution.
\end{abstract}


\section{Introduction}

Recent infrared observations suggest that spiral galaxies may exhibit a
greater fraction of bars than was once inferred from (nuclear saturated)
optical data (Block \& Wainscoat 1991, Zaritsky, Rix \& Rieke 1993). Optical
surveys already reveal that about two-thirds
of disk galaxies have a bar (Sellwood \& Wilkinson 1993, Martin 1995).
Indeed, it is believed that most spirals have probably harbored a
self-destructive
bar at one time or another during their evolution (Friedli \& Benz 1993,
hereafter FB93;
Friedli \etal 1994, von~Linden 1996).
Such bars are easily formed through global instabilities of a rotationally
supported disk, or via interactions, and gas redistribution by the bar can
cause
its own dissolution. Box- or peanut-shaped bulges also provide evidence for the
current existence of a bar in edge-on systems. Once formed, such bars
or oval distortions are believed to be efficient carriers of disk material
into the central regions through angular momentum torques and viscous transport
(Kormendy 1993, hereafter K93; Pfenniger 1993, hereafter P93; Martinet 1995,
hereafter M95, and references therein).  Viscous dissipation has been proposed
by many (Silk \& Norman 1981, Lin \& Pringle 1987, Yoshii \& Sommer-Larsen
1989,
Saio \& Yoshii 1990, Struck-Marcell 1991, hereafter SM91) to explain the
exponential
distribution of the
stars in galactic disks.  An exponential profile in the central regions is also
expected from non-axisymmetric disturbances which will induce inward radial
flow
of disk material (Combes \etal 1990, Pfenniger \& Friedli 1991, SM91, P93).

While common use of the $r^{1/4}$ law (de Vaucouleurs 1948) for fitting the
light
profile of disk central regions has endured unprecedented popularity over the
years, evidence for exponential fall-off of the central light distribution
(or at least departures from $r^{1/4}$) in disk galaxies is not recent
(van~Houten 1961, Frankston \& Schild 1976, Kormendy \& Bruzual 1978, Burstein
1979,
Shaw \& Gilmore 1989, Kent \etal 1991, Andredakis \& Sanders 1994). Recently,
Andredakis, Peletier \& Balcells (1995, hereafter APB) used the generalized
exponential law of S\'ersic (1968)
with their images for 30 early-type spirals and data from Kent (1986) for 21
late-type systems, to show that bulge profiles vary with Hubble types.
S\'ersic proposed the form

$$\Sigma(r) = \Sigma_\circ exp\{-(r/r_\circ)^{1/n}\} $$

{\noindent where $\Sigma_\circ$ is the central surface brightness (CSB),
$r_\circ$ a
scaling radius, and $n$ is the exponent variable. If $n=1$, one has a pure
exponential profile with $r_\circ$ as the scale length; with $n=4$, one
recovers
a deVaucouleurs profile.}

In this Letter, we use the red and infrared light profiles for 326 spirals
to confirm the result by APB that central regions of late-type spirals
are best fit by an exponential and separately show that the bulge and disk
scale
lengths are closely coupled.  Unlike APB, we interpret these
results as support for a picture of secular evolution in disk galaxies.

\section{Method}

We have combined the large collection of deep $r$-band profiles of Sb/Sc
galaxies
by Courteau (1992, 1996; hereafter C96) and BVRIHK photometry by de~Jong (1995,
hereafter dJ95) for face-on Sa--Sm galaxies. Courteau's sample was selected for
Tully-Fisher mapping of peculiar velocities and thus includes galaxies of
moderate
to high inclinations ($50\arcdeg < i < 80\arcdeg$). The original collection
counts
350 spirals but 290 were kept for final decompositions; the others were too
small to
successfully resolve the bulge.
Sky subtraction errors, probably the greatest source of uncertainty
in B/D decompositions, were investigated and treated for with great care (C96).
de~Jong's sample was constructed to examine the global structure of spiral
galaxies
including the elusive law of constant disk CSB (Freeman 1970, dJ95,
McGaugh \etal 1995, C96), and study the observable effects of dust and stellar
populations in spirals. The IR data is extremely
useful for studying the central light, unobscured by the dust; profile
decomposition
is thus less likely to suffer from internal absorption effects (Phillips \etal
1991). Moreover, the
wide sampling of Hubble types allows one to examine morphology correlations.
Both samples were selected from the UGC (Nilson 1973) and only include
systems with normal-looking appearances and clean stellar foreground.
The typical galaxy size is 2\farcm3.
Many of these galaxies were classified as non-barred by Nilson but direct
examination of the IR image or removal of a smooth bulge and disk fit
from the optical image often reveals the presence of a bar. About a third
of de~Jong's galaxies required a bar as extra component to fit the surface
brightness
distribution.

Bulge-to-disk (B/D) decompositions were done independently by de~Jong (dJ95,
1996)
and Broeils \& Courteau (1996, hereafter BC96). We combine our results for the
benefit of
this Letter.  Unfortunately, no overlap exists between the two catalogs
to test for systematic errors. dJ95 used both major-axis profiles (1D) and full
image (2D) B/D
decompositions of his thesis sample with $n=1$, 2, and 4 for the central
regions and a standard disk exponential. 2D decompositions offer the advantage
of fitting for any additional central component, like a bar, ring or lens. They
also
yield smaller error bars than 1D on each of the fitted parameters, and allow
for a
more robust recovery of simulated input parameters (Byun \& Freeman 1995,
dJ95).
Still, we find that our results do not depend strongly on the type of
techniques adopted.

BC96 decompose the light profile of Courteau's galaxies
as bulges with $n=1$ and 4, plus an exponential disk. This 1D technique
is similar to that introduced by Kormendy (1977) which consists in $\chi^2$
fitting the bulge and disk profiles simultaneously via non-linear least-squares
analysis.  2D decompositions for the same galaxies, in a study of correlations
with
the rotation curve, will be presented elsewhere (Courteau \& Broeils 1996).
While dust is more conspicuous at $r$ than $K$ for central regions, we believe
that
our results are not significantly affected by dust since they statistically
reproduce the same range of values for de~Jong's galaxies in
the $K$-band.  In all cases, seeing is accounted for by convolving the model
profiles
with a Gaussian PSF with a dispersion measured from field stars.

Since bulges of late-type spirals are small and can be affected by
atmospheric blur, we tested our decomposition routines with thousands of
artificial
images (or surface brightness profiles in 1D) with a wide range of input
parameters
and various values of $n$ to derive a space of recoverable parameters (see also
Schombert \& Bothun 1987, Byun \& Freeman 1995, and APB).  The simulated images
were
convolved with a seeing Gaussian and include photon and readout noise.
Our results suggest that all values of $r_b$ above 1\arcsec\ are successfully
recovered for all observed seeing values, as are the values between 0\farcs5
and
1\farcs0 with a seeing disk below 2\arcsec.
Values of $r_b < 0\farcs5-0\farcs6$ are not reliable unless seeing is better
than 1\farcs5. In light of these results, we reduced Courteau's sample to 243
galaxies. The measuring uncertainty in CSBs and scale lengths is dominated by
sky subtraction errors. Full details are given in dJ95 and BC96.

\section{Results}

{}From examination of the reduced $\chi^2$ values, we find that most late-type
spirals are best fitted by a double-exponential.
60\% of de~Jong's total sample, which includes most of the Sb and Sc galaxies
and all galaxies
of later type, is best modeled with the double-exponential.  A quarter of the
sample,
mostly Sa-Sab's, is best
modeled with a $n=2$ bulge.  Few galaxies ($\sim 15$\%) are more appropriately
fit by a deVaucouleurs law in the central regions.
BC96 find that about 85\% of their Sb and Sc's are best fit by the
double-exponential
while the remainder is better fit with an $r^{1/4}$ bulge profile.
Granted that $n=1$ for most late-type spirals, we adopt exponential profiles
for
the central region and disk of all galaxies in our sample and compute their
scale length
ratio\footnote{The results are unchanged if we restrict ourselves to the
subsample
of pure $n=1$ central profiles only.}. Figures 1 and 2
show the measured scale lengths for our divided sample in the $r$-band. The
inset histograms
show the range of possible values for the scale lengths ratio.
Displayed in physical units, the correlation would show a slightly
narrower stretch from 0 to 10 kpc (dJ95).  Angular units show that
the correlation is not affected by resolution effects.

Combining both $r$-band data sets, we find
$ r_b/r_d = 0.08 \pm 0.05 $ while de~Jong galaxies alone at $K$\arcmin\ yield
$ r_b/r_d = 0.09 \pm 0.04 $ (not shown). Hence the effect of dust is not
alarming.
Results at $B$ or $V$ are not presented because the scatter becomes too large.
Note also that this result would remain unnoticed were one to fit
(inappropriately)
all the inner regions with an $r^{1/4}$ law.  The histogram peak value may not
be
determined with great certainty (2$\sigma$) but the importance of Figures 1 \&
2 is
the demonstration of a restricted range for the scale length ratio.
Earlier-type spirals
(Sa's$-$Sab's) in Figure 2
appear systematically below the nominal line as a result of ``improper''
fitting of
an $n=1$ bulge profile. Nonetheless, the discrepancy is not large and
early-type
spirals closely follow the overall trend suggesting that the Hubble sequence of
spirals
is scale free.
While each Hubble type comes with a range
of different diameters and total luminosities, the constant ratio of B/D scale
lengths
appears to be independent of galaxy type.  Disks of larger scale lengths would
simply
form bulges with correspondingly larger scale lengths.

\section{Discussion}

A correlation between B/D scale lengths is best understood in a model where the
disk forms first and the bulge that naturally emerges is closely coupled to the
disk.  Were the bulge to form first, it would be hard to understand how a small
dynamically hot component could directly influence the disk global structure.
In standard cosmological models, galaxy formation occurs when matter decouples
from the uniform Hubble flow and a disk (for spirals) is formed by
dissipational
collapse of the initial gas.
The observations presented here concern the stellar component of central
galactic
regions and the main obstacle in making the bridge between the phase of
formation
and current observations is the limited numerical resolution and ad hoc
formulation
that bulge stars form from primordial high-$\sigma$ fluctuations (Katz 1992,
Navarro
and White 1993, Steinmetz \& M\"{u}ller 1994).
The secular evolution models in which the bulge is formed naturally from the
disk
have more predictive power. In a model of viscous evolution, an exponential
disk
emerges from the redistribution of angular momentum by the viscosity of the
gas, if
star formation occurs on roughly the same time-scale (Lin \& Pringle 1987,
Yoshii \&
Sommer-Larsen 1989, SM91).
Such a model will automatically develop a ``bulge'' whose properties only
depend on
the relative time-scales of star formation and viscous transport and on the
total
angular momentum (Combes \etal 1990, Saio \& Yoshii 1990, SM91), and will
produce correlated B/D scale lengths.
Efficiency of transport will be improved with a bar or oval distortion which
can be
triggered by a global dynamical instability in the disk or induced by
interactions with a satellite (M95). This in turn, will induce funneling
of disk material into the central regions. Disk stars will be heated vertically
up to 1--2 kpc above the plane by resonant scattering off of the bar-forming
instability
and a ``bulge-like'' component with a nearly exponential profile will emerge
due to
relaxation induced by the bar (K93, P93, and references therein).
Secular accumulation or satellite accretion of only 1-3\% of
the total stellar disk mass near the center is sufficient to induce dissolution
of
the bar into a spheroidal component (Pfenniger \& Norman 1990, Friedli 1994,
M95).
Secular evolution with bar transport can thus be a viable explanation for bulge
formation
in late-type spirals whether a bar is or is no longer present.

Our measurements of exponential stellar density profiles as well as a
restricted
range of B/D scale lengths provide strong observational support for secular
evolution
models.  Self-consistent numerical simulations of disk galaxies evolve toward a
double-exponential profile with a typical ratio between bulge and disk scale
lengths
near 0.1 (Friedli 1995, private communication) in excellent agreement with our
measured values.  Circumstantial
evidence also include the colour of bulges which is known to change only slowly
from the inner disk into the central regions (Courteau \& Holtzman 1995, 1996;
dJ95, Peletier \& Balcells 1995) and the fact that, unlike ellipticals,
bulges' kinetic energy comes mostly in rotation which must be imparted
from the disk (Kormendy 1985, P93).
Subtraction of an elliptical profile fit from the original
galaxy image shows residual spiral structure that extends all the way into the
center of the galaxy for the majority of Courteau's galaxies.
This provides further support (though not absolute evidence) for an association
between the bulge and disk (K93, Zaritzky, Rix \& Rieke 1993).

Note that APB reject secular evolution on the basis of their continuous
spectrum
of the index $n$ versus morphological type. They propose that the smooth
sequence
they observe (see their Figure 5) can only result from a single mechanism of
bulge
formation. Given the large scatter in their diagram, such a conclusion seems
ill-based.
A more plausible scenario in which big bulges ($>2$ kpc as in Sa's) form
principally
from a minor merger and smaller bulges ($1-2$ kpc; Sd's $\rightarrow$ Sab's)
form mainly via secular evolution is not likely to leave any bi-modal imprint
on the spectrum of ``n'' as both processes will operate to some degree of
efficiency
for all Hubble types\footnote{The observed scatter in both APB's Figure 5
and our Figs. 1 \& 2 is probably correlated with some third parameter that
could provide
a link between secular evolution and merging scenarios. This will be
investigated elsewhere.}
(P93, M95).

Chemical evolution is also likely to be affected by gas flows in the center of
galaxies. For galaxies undergoing central star formation bursts, Friedli \etal
1994
show that the gas-phase abundance gradient should be characterized by two
separate
slopes corresponding to the inner and outer regions of the disk. We expect that
the
break point will correlate with the transition between the central and outer
disk
exponentials.  This remains to be tested.  A metallicity-velocity dispersion
relation
for the core is expected as well, though current nuclear stellar abundances are
too
uncertain to provide conclusive evidence (Friedli \& Benz 1995). Boroson (1980)
measured central metallicities (Mg2 index) in the bulge of 24 spirals and found
a higher correlation with bulge light than with total light though with poor
statistics (see also Jablonka \etal 1995). Such a picture, if true, would
suggest
a decoupling of the bulge and disk at formation, in stark contrast with models
of secular evolution.
Finally, the possible existence of an intermediate age population with
some ongoing star formation in the bulges of our Galaxy, M31, and M32
(Davies \etal 1992; Rich \etal 1993; see also Renzini 1993) would lend
support to secular evolution models.

Tests for secular evolution will greatly benefit from the measurement of
high-resolution central abundances and metallicity gradients coupled with
studies of the 2D infrared light distribution to fit the exponent $n$ for a
large sample of Sd to Sa's. Nearby and more distant samples are needed to
test for the effects of galaxy evolution.

\acknowledgments

We are grateful to P. Martin for useful conversations and suggesting new
references. Daniel Friedli also provided numerical tests and valuable comments
on the first draft of this paper.

\clearpage

\begin{figure}
\figurenum{1}
\plotfiddle{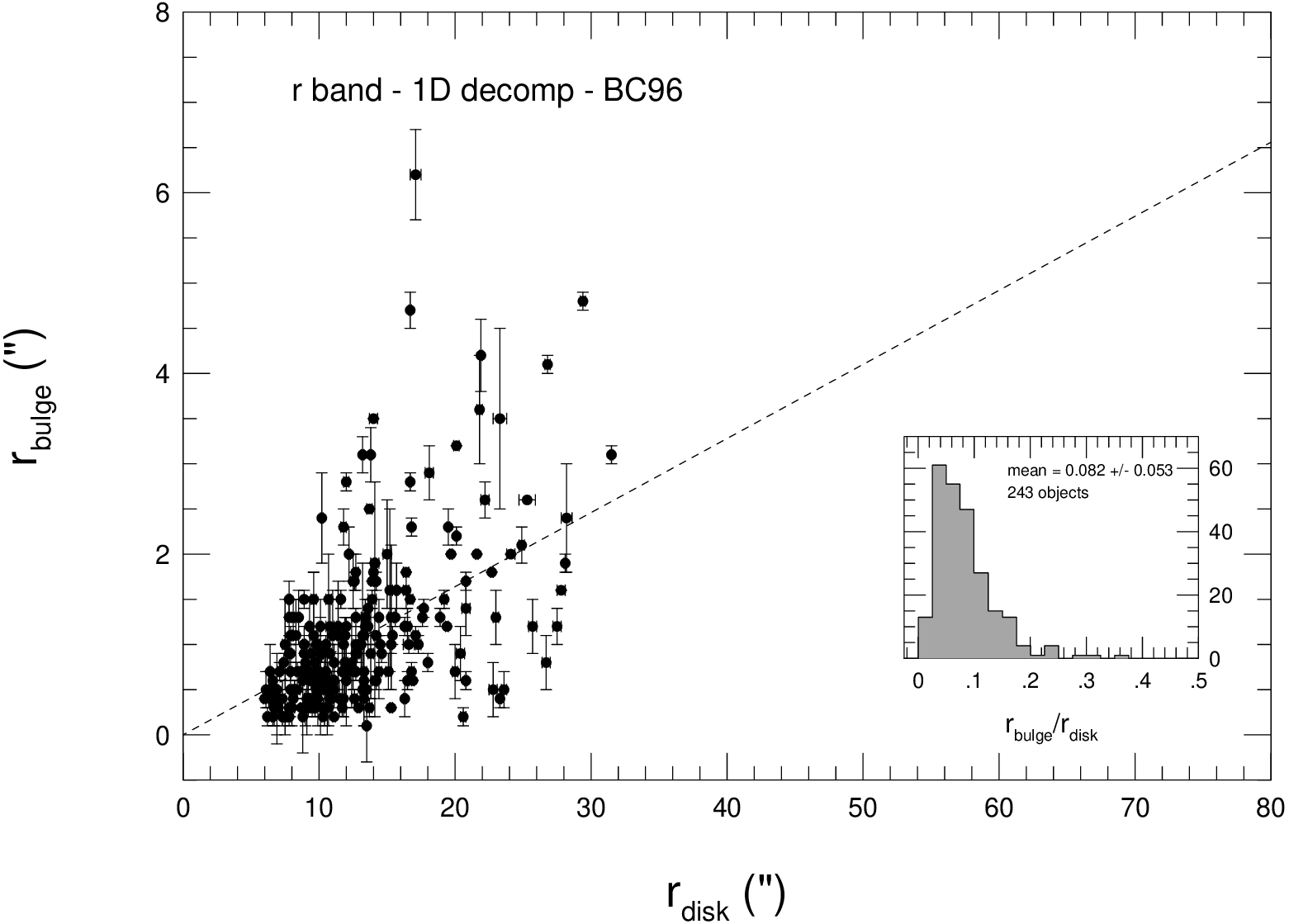}{4.5in}{0}{70}{70}{-285}{0}
\caption{Fitted scale lengths for the bulge and disk in arcseconds from BC96
 for 243 Sb \& Sc galaxies in Courteau's sample (C96). The dashed line
 is a fit to the data; its slope is given by the mean ratio of
 $r_{\rm bulge}/r_{\rm disk}$.
 The histogram shows the distribution of B/D ratio (unweighted by the errors).}
\end{figure}

\clearpage

\begin{figure}
\figurenum{2}
\plotfiddle{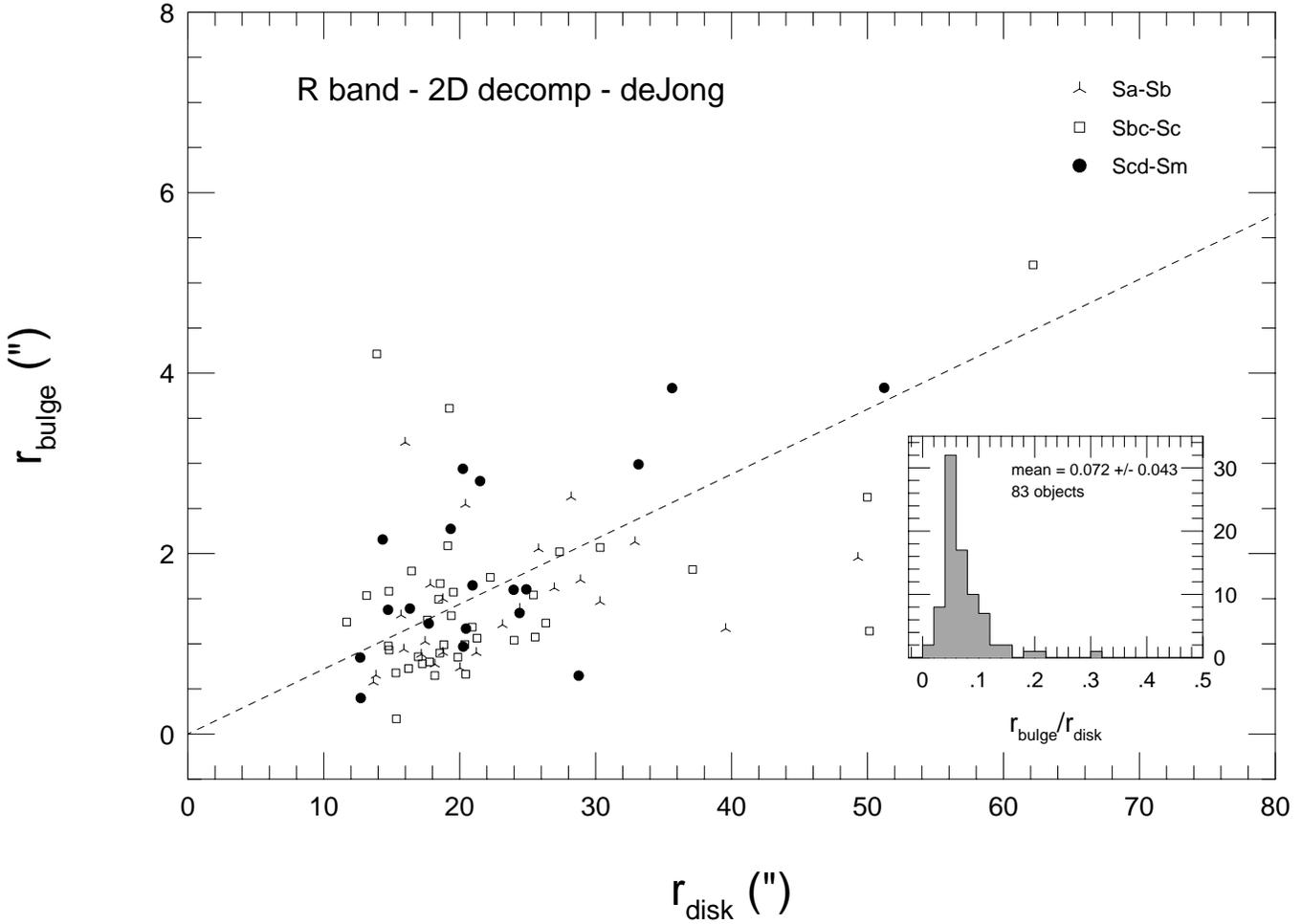}{4.5in}{0}{70}{70}{-285}{0}
\caption{Same as Figure 1 for de~Jong's galaxies. The different symbols
 represent different ranges of Hubble types.  The formal fit error bars
 for these 2D decompositions (not including sky uncertainties) are
 comparable to the size of each point and were not plotted to preserve
 clarity.}
\end{figure}

\end{document}